# Dynamic Buffer Sizing for Out-of-Order Event Compensation for Time-Sensitive Applications

WOLFGANG WEISS, JOANNEUM RESEARCH Forschungsgesellschaft mbH, Austria
VÍCTOR J. EXPÓSITO JIMÉNEZ, VIRTUAL VEHICLE Research GmbH, Austria
HERWIG ZEINER, JOANNEUM RESEARCH Forschungsgesellschaft mbH, Austria

Today's sensor network implementations often comprise various types of nodes connected with different types of networks. These and various other aspects influence the delay of transmitting data and therefore of out-of-order data occurrences. This turns into a crucial problem in time-sensitive applications where data must be processed promptly and decisions must be reliable.

In this paper, we were researching dynamic buffer sizing algorithms for multiple, distributed and independent sources, which reorder event streams, thus enabling subsequent time-sensitive applications to work correctly. To be able to evaluate such algorithms, we had to record datasets first. Five novel dynamic buffer sizing algorithms were implemented and compared to state-of-the-art approaches in this domain. The evaluation has shown that the use of a dynamic time-out buffering method is preferable over a static buffer. The higher the variation of the network or other influences in the environment, the more necessary it becomes to use an algorithm which dynamically adapts its buffer size. These algorithms are universally applicable, easy to integrate in existing architectures, and particularly interesting for time-sensitive applications. Dynamic time-out buffering is still a trade-off between reaction time and out-of-order event compensation.

This research was funded by the Federal Ministry for Climate Action, Environment, Energy, Mobility, Innovation and Technology (BMK) within the framework of a sponsorship under the projects "Collaborative Robotics" (CollRob) and "Multi-Dimensional Sensor Data Time Series Analysis" (MUST). This research was partly funded by the Comet Project "Dependable, secure and time-aware sensor networks" (DeSSnet). DeSSnet is funded within the context of COMET - Competence Centers for Excellent Technologies by the Federal Ministry for Climate Action, Environment, Energy, Mobility, Innovation and Technology (BMK), the Federal Ministry for Digital and Economic Affairs (BMDW), and the federal states of Styria and Carinthia. The programme is conducted by the Austrian Research Promotion Agency (FFG). Research leading to these results has received funding from the EU ECSEL Joint Undertaking under grant agreement no. 737459 (project Productive4.0) and from the partners' national funding authorities FFG on behalf of the Federal Ministry for Climate Action, Environment, Energy, Mobility, Innovation and Technology (BMK) and the Federal Ministry of Education, Science and Research (BMBWF). Moreover, the authors would like to acknowledge the financial support of the COMET K2 – Competence Centers for Excellent Technologies Programme of the Federal Ministry for Climate Action, Environment, Energy, Mobility, Innovation and Technology (BMK), the Federal Ministry for Digital and Economic Affairs (BMDW), the Austrian Research Promotion Agency (FFG), the Province of Styria and the Styrian Business Promotion Agency (SFG). An earlier version of this article [23] appeared in the Proceedings of the 2nd International Conference on Internet of Things, Big Data and Security 2017.
Authors' addresses: Wolfgang Weiss, JOANNEUM RESEARCH Forschungsgesellschaft mbH, Steyrergasse 17, Graz, 8010, Austria, wolfgang.weiss@joanneum.at; Víctor J. Expósito Jiménez, VIRTUAL VEHICLE Research GmbH, Inffeldgasse 21a, Graz, 8010, Austria, victor.expositojimenez@v2c2.at; Herwig Zeiner, JOANNEUM RESEARCH Forschungsgesellschaft mbH, Steyrergasse 17, Graz, 8010, Austria, herwig.zeiner@joanneum.at.

## 1 INTRODUCTION

Internet of Things applications are on the rise as they accelerate digital transformations many industries may benefit from. As a result, the number of connected devices and Internet of Things networks increase with the consequence that also the volume of generated data increases. In addition, end users expect timely and accurate results from connected applications.

Technically, sensor networks and Internet of Things applications can be built upon multiple, independent, and distributed nodes generating and processing data. In applications used today it is no longer sufficient to store data and retroactively process it. Data needs to be processed online and decisions must be made near real-time. To be able to do so, a processing agent, such as an event processing engine, collects the event streams from different sources and processes them. This raises a couple of issues, mainly introduced by various delays, e.g. when detecting events, transferring events to its destinations, or processing events. Suppose you are searching for a pattern of event $A$ being followed by event $B$, with each event coming from a different source in a distributed system. This requires that all delays are either zero or of constant length, otherwise it is likely that out-of-order events occur. These are events which arrive too late, e.g. in our example event $A$ occurred before event $B$ but arrived after event $B$ in the event processing engine. Event processing is order and time-sensitive and therefore assumes temporally correct ordered event streams to be able to create correct results.

One prominent mechanism for reordering events is to use time-out buffering. Thereby, events are delayed in a buffer until they reach a preset time-out. Events arriving after the time-out cannot be reordered. As the transmission time of events varies (like the travelling time of cars on a highway at the rush hour and off-peak times), it gets necessary to dynamically adapt the buffer size accordingly. This keeps the buffer time overall on a useful lower level, while relieving the system engineer who designs and implements sensor network applications from the burden to exactly preconfigure a time-out buffer. Instead, the engineer just needs to configure basic parameters in which the algorithms operate. Dynamic buffer sizing algorithms calculate the buffer time from the incoming event stream. Depending on the sizing strategy of the algorithm and the chosen parameters, the resulting buffer time is usually higher than the minimum necessary (subsequently referred to as *"overfitting buffer time"* ), which has negative effects on the reaction time. As there occur a number of unpredictable outliers of delayed events, and also the desire to keep the buffer time low, it can still happen that not all out-of-order events can be compensated (subsequently referred to as *"not compensated events"* ). This also defines the trade-off space in which these algorithms operate.

We evaluated dynamic buffer sizing algorithms for multiple and distributed sources, which reorder event streams, enabling subsequent time-sensitive applications to work correctly. These buffer algorithms are able to deal with varying temporal delays. After the taxonomy of [14], this project can be classified as a data related fusion aspect dealing with data inconsistency, where the data inconsistency refers to as disordered data. In the domain of sensor measurement fusion this is often called *"out-of-sequence measurements"*. We stay in the domain of event based systems and event processing, where the literature refers to this problem as *"out-of-order events"* (cf. [8, 9, 19]).

### 1.1 Outline

In this paper we want to find answers to the following questions: Is there a dynamic time-out buffering method which is preferred over a static buffer? Is a dynamic time-out buffering method applicable to fuse data from multiple distributed and independent sources for time-sensitive applications? The expectations for a dynamic buffer are high. The buffer should be as small as possible while ensuring that all incoming events can be fully reordered. The buffer should adapt itself to environmental conditions and changes of these conditions, such as varying transmission delays or other influences.

We build upon previously published results [23] and extend it in several ways: More datasets were recorded and a deeper analysis of the datasets is provided here. The algorithms were improved and two new algorithms were added. A much deeper analysis of the algorithms is provided in this project, where all datasets were applied.

Based on this project we are making following contributions:

(1) *Five novel buffer sizing algorithms:* novel algorithms have been implemented which dynam- ically calculate the buffer time through derived properties of the current environmental situation.
(2) *Datasets for the evaluation of out-of-order compensation algorithms:* 21 datasets containing more than 500,000 events were recorded. These datasets are suitable for the evaluation of out-of-order event compensation algorithms. All datasets were made open source under a Creative Commons license.
(3) *Evaluation of the algorithms:* a detailed evaluation was carried out to compare the implemented algorithms by using the open source datasets.

The target audience of this paper are algorithm researchers in the field of distributed computing, and engineers in the field of sensor networks and Internet of Things applications. They benefit as follows:

(1) Researchers who design and implement algorithms for out-of-order event compensation. They benefit from the existing algorithms and their evaluation. They are free to use the datasets to evaluate their implementations.
(2) Engineers who design and implement sensor networks or Internet of Things systems which have a time-sensitive aspect in their application. They benefit from the description of the problem and its influencing factors. They know which temporal properties need to be considered. They can select a suitable algorithm for their use case and know how to configure it.

The next section gives an introduction on the background of this domain and lists the requirements. An overview of existing and related work in the field of out-of-order event processing and compensation is given in Section3. The content of the dataset for evaluating the buffer algorithms dealing with out-of-order events is discussed in Section4. Section5describes the buffer algorithms, and the results of the evaluation is discussed in detail in Section6. A reflection of the results and the initial research questions is given in Section7and, finally, Section8concludes this paper.

## 2 BACKGROUND

Subsequently we give an introduction on the technical background of this problem domain. Afterwards we list two use cases where out-of-order events occur and where temporally correct ordered event streams are essential to guarantee correct results. Finally, this leads to a list of requirements which must be fulfilled by the algorithms and the recorded datasets.

### 2.1 Technical Background

The distributed nature of sensor networks and Internet of Things applications raises a couple of problems which were already investigated in general in distributed computing research (cf. [5]). Important aspects here are that there is no global clock, and that the program execution is concurrent.

No global clock means that there is not a single global notion of correct time, but a shared idea of time is a prerequisite for temporally close coordination between nodes. Local clocks of nodes can be synchronized but inaccuracies remain.

Figure 1 illustrates a generic architecture of a distributed and event based system, which also represents the reference architecture of this project. This system contains various distributed components in spatially separated locations. *Sensors* are the devices which analyse its environment, and *sensor processing nodes* process the raw sensor data and deliver enriched data to its consumers through the connected computer network. Sensor processing nodes are intelligent devices which store their state, and have a synchronized clock. Consumers of event data can be end user devices, or other processing nodes which further combine and process information. Sensors at a remote place or on a mobile platform may not be able to be permanently online. Therefore they have a *sensor cache* which forwards the events whenever the network or energy level allows it. This architecture also reflects new computing paradigms such as fog / edge computing which employs data processing near the device or user, and no longer only centralized in the cloud [26].

An *event stream* is referred to a set of associated events which might be temporally totally ordered, meaning there is a well-defined timestamp order within the stream. An *event processing agent* is a software component which collects and processes the events. As a result it emits *derived events* based on the processing logic. An *event pattern* is a template which defines one or more event combinations, e.g. *A not followed by B within two time units* (cf. [8]).

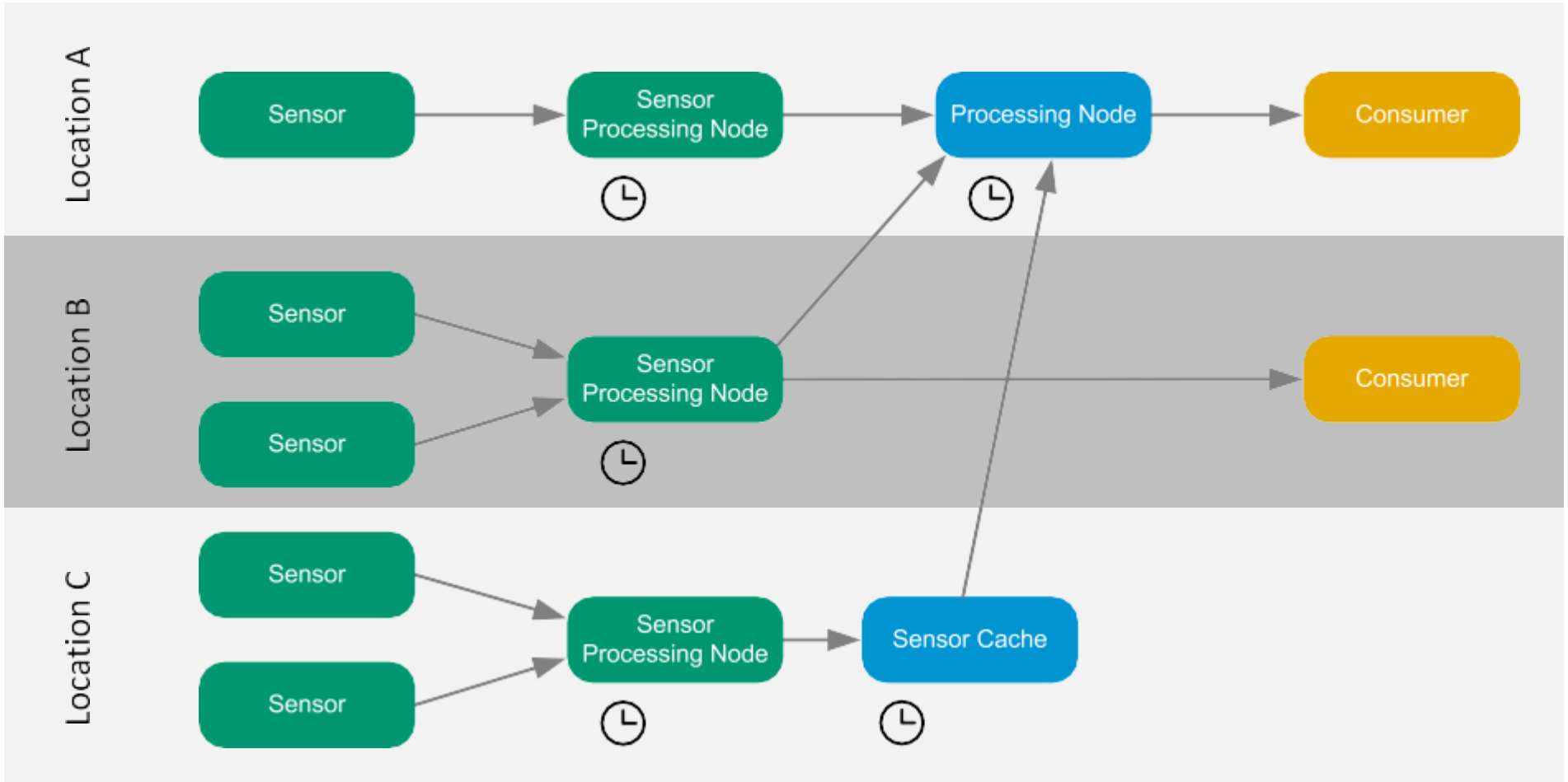

Fig. 1. The reference architecture of this project which is event-driven and distributed. Sensor processing nodes do the processing of raw sensor data and deliver enriched data over the network. Sensor caches are optionally deployed for remote or mobile platforms. All data processing nodes have their own synchronized clock.

With the possibility of being always online and the need to get instant results and notifications, it has become necessary to process data timely. One option is to use an event processing engine. *Event processing* (or complex event processing) is the computation that performs operations on events from possibly various sources. An important fact is that event processing is order and time-sensitive and therefore assumes temporally correct ordered event streams to be able to create correct results. Consequences of not correctly ordered event streams are:

- *False negative:* no event detected when an event should have been detected.

- *False positive:* detected an event when no event should have been detected.
- *Wrong calculations:* a wrong value is calculated when using a temporal sliding window.

An example: an event processing agent searches for matches of the event pattern *"A followed by B followed by C within 3 time units"* in an event stream. If the input stream has out-of-order events, as illustrated in figure2, then the event processing agent will not be able to match the event pattern defined above. In this case, event *C* with timestamp 13 and event *A* with timestamp 16 arrived too late. After reordering the event stream based on the assigned timestamps (see figure3), the event processing agent is able to match the event pattern and create derived events.

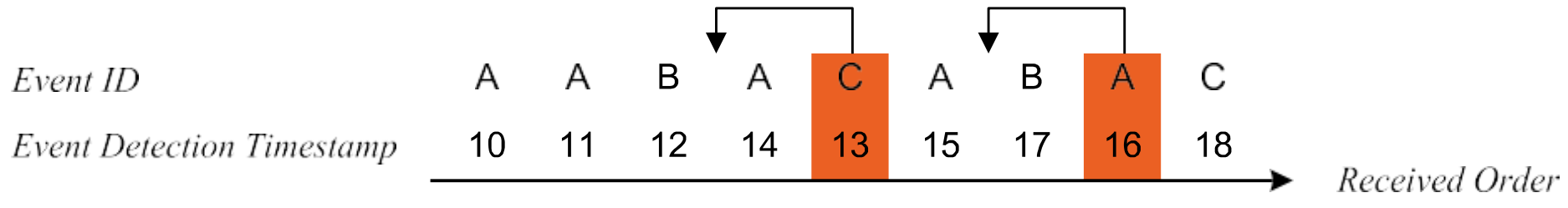

Fig. 2. An event stream containing out-of-order events (C-13 and A-16).

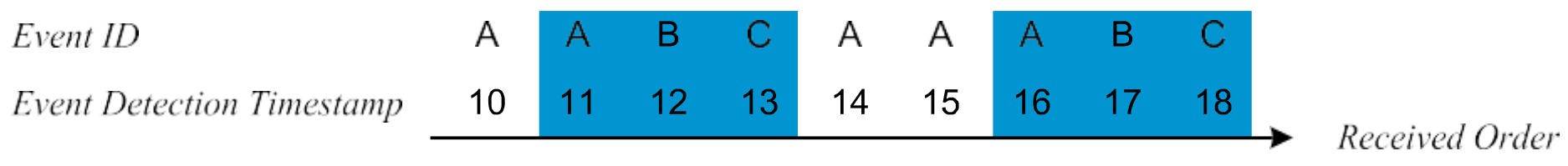

Fig. 3. The correctly reordered event stream allowing the detection of the pattern *"A followed by B followed by C within 3 time units"*.

## 2.2 Use Cases

Two use cases from different domains are presented where temporally correct ordered event streams are essential to guarantee correct results. In the first use case, we illustrate an example to monitor an industry facility based on a sensor network. A central processing node autonomously makes decisions to stop the facility in the case of a failure. In the second use case, a connected car is informed by other vehicles and has to decide on which lane it should drive in order to avoid an upcoming obstacle.

The first use case utilizes a sensor network to monitor an industrial facility to detect a leakage of pressurized air supplies. Leakage is typically 10-20% of the supply on an annual average, even in well-maintained systems. The sensors in this example measure the pressure, temperature and vibration of pipelines and other system facilities. Early detection of air leakages can save electrical energy which is used for the pressurized air generation. The sensors are spread over the area of the industry plant and are connected for cost reasons via a wireless network. A central processing node receives the data from the distributed nodes, analyses it, and decides if a leakage happened and where it is. As there are many interferences in this kind of harsh environment, it is possible that event data is affected by some delays when transmitted from the sensor to the fusion centre.

We can assume that the system is in normal operation mode when the pressure value reported by sensor A (mounted at the beginning of the system) corresponds with the pressure value reported by sensor B (mounted somewhere else in the system). Meaning, when the pressure in sensor A increases, then the pressure in sensor B must also increase. If these values do not correspond, we can assume that there is a leakage somewhere and the system must be maintained. In the case of an out-of-order event from sensor B, a false positive would be detected and a wrong decision would be taken at the fusion centre.

The second use case is based on the domain of connected cars. Imagine there is a multi-lane road with two cars driving side by side in the same direction. The cars scan the road in front of them and report important observations to the cars behind them over a wireless ad hoc network. The car in the right lane in front of the convoy recognizes a person on the street who is 200 meters in front of it. At a speed of 80 km/h this will not require any immediate action, but this occurrence should be reported to the other cars by alerting them: *Event 1 (car 1, right lane, person on the lane, 200m)*. The leading car on the left lane receives this event and reports that its lane is free for the next 300 meters: *Event 2 (car 2, left lane, lane is clear, 300m)*. The person moves on, causing the cars to report the following events: *Event 3 (car 1, right lane, person on the lane, 175m)*, and another report by the car on the left lane: *Event 4 (car 2, left lane, lane is clear, 300m)*. The person reaches the left lane: *Event 5 (car 2, left lane, person on the lane, 150m)*. Now, the right lane is free: *Event 6 (car 1, right lane, lane is clear, 300m)*.

The vehicles following car 1 and car 2 must ensure to process these events timely and in the correct temporal order to react fast and adequately. If the events are in correct temporal order, they can infer the following information: The right lane is clear for the next 300 meters, while on the left lane there is a person in a distance of 150 meters, and there is a person moving from right to left. This example illustrates a use case where the order of events is of significant importance and where a minimum delay is required for further actions. This use case also has multiple event producers which work independently of each other.

### 2.3 Requirements

Derived from the above described use cases and the technical environment where those systems operate, we can specify the following requirements for out-of-order event compensation algorithms:

- Deal with *varying transmission delays*: nodes can be deployed on various network types such as wired, wireless, or mobile networks. This in turn can be a local network or the public internet. All of this influences the transmission time of data packets from the origin to their destination.
- Work with *varying event frequency*: events can be produced periodically but also sporadically when a certain condition has occurred.
- *Independent nodes*: nodes in an Internet of Things application may have no awareness of each other.

## 3 RELATED WORK

Our research is focused on detecting and compensating temporal out-of-order occurrences of event data in distributed systems. In practice it is a combination of different topics where someone has to deal with the characteristics of distributed systems, such as network delays, clocks and partial system failures. In this context, it is also a method for multi-sensor data fusion [14]. The problems which arise when dealing with events in time and space was recently discussed in [11]. The author lists typical influencing factors why delays occur. As an alternative solution he suggest that applications should be able to deal with inaccurate data.

Out-of-order compensation can be solved by using different handling approaches. For example, buffer-based approaches in which a buffer is used to sort the incoming events by their timestamp or sequence ID. Following this approach, the authors in [19] use a K-slack buffer approach where the buffer length (K) is continuously recalculated and adjusted according to the amount of detected out-of-order events and the delay of these events. This system does not use a local or global clock but instead derives the current time from incoming events. The buffer size (the K value) is calculated by analysing all previous delay measurements and adding a safety margin on top of it which is calculated by the standard deviation and pre-defined scaling factor. Omitting a dedicated clock and the derivation of the current time has a considerable disadvantage as the event transmission has a delay on its own, which may also depend on the payload of the event itself. In our approach we rely on a dedicated clock synchronization mechanism. However, their approach has given us a basis to design some of our algorithms, but with improvements on the buffer size calculation. In [21], the same authors extend their work on low-latency constraint systems and look at the question of how out-of-order events can be compensated by using the different delays between hosts in distributed systems, thereby choosing the best route of compensating for the delays to guarantee the correct event order. Another approach is given in [17] in which the authors handle the out-of-order problems by using parallel query models. In this case, the approach also introduces different dropping ratios. According to the results, the latency can be reduced in more than one order of magnitude with a 1% dropping ratio in comparison with the K-slack approach. Unfortunately, our research is focused on keeping as many events as possible and the usage of dropping ratios to improve the latency is not taken into account.

An Adaptive, Quality-drive K-slack (AQ-K-slack) is another K-Slack based approach that is shown along the publications [12, 13] in which Ji et al. are able to reduce the latency more than 50% compared to the standard K-slack approach. In contrast to our approach that is generally applicable and independent of the successive computation logic in which the distribution of the transmission delays are investigated to calculate the buffer size, the authors introduce an error model for query results in which the window coverage metric is bounded to the windowing function.

Although our research is focused on buffer-based approaches, there are two more common ways to handle this issue: speculation-based and punctuation-based. Speculation-base uses revision techniques when an out-of-order event is reached to recalculate the previous handled events affected by this out-of-order event. This kind of approach may be highly CPU intensive especially when many out-of-order events are detected due to this revision technique. A good example is given in [20] in which the authors combine speculation and buffer-based in their solution. On the other hand, in a punctuation-based [4, 15] approach, special tuples are included in the data stream to determine if an out-of-order event is reached. For example, Ji et al. in [15] include a timer-driven punctuation which is propagated by the query operators and used to monitor query latency and detect operator failure.

Another method is given in [2, 9] in which the authors developed an open source system for complex event processing called ETALIS[1]. They incorporate an out-of-order semantic that is used to describe the event pattern as a set of rules which allow to detect and compensate the out-of-order event. According to the evaluation, it is able to improve the memory consumption in cases of high out-of-order event rates and heavy computing. Unlike our work, this out-of-order events compensation mechanism is integrated in the complex event processing engine and not an independent component as we desire. The articles [16, 25] introduce a new method to handle out-of-order events. They use a new kind of structure that not only stores the current instance status, but also the previous one. The algorithm stores prevent instances until the amount of current event time unit, window length and K length is less than the highest time unit received. If this happens, the system will be able to safely purge this event. The K value (the buffer size) is a static and predefined constant. In our project, we don't just store the previous event but all events in a limited time window which allows us to dynamically adapt the buffer size.

[1]https://code.google.com/archive/p/etalis/

A prerequisite to deal with time in a distributed system is to agree on a common notion of time, and here clock synchronization comes into play. One of the simplest ways to synchronize clocks is Christian's algorithm [6] which is basically a request/response method that depends on a single server and the request round-trip time. The Network Time Protocol (NTP) [18] is well known and widely used. It synchronizes the clocks in a predefined interval, but clocks may drift on their own, and the synchronization process has some inaccuracies too. A more advanced variant is the Precision Time Protocol (PTP) [1, 7]. It is made to provide high precision, but it is limited for local networks with a few subnetworks. There are also other variants which do not need a central server, such as vector clocks [22]. Thereby all nodes send the clock vector to each other to calculate the current time.

## 4 DATASETS

This section introduces the datasets which have been recorded and generated with the purpose to evaluate out-of-order event compensation algorithms. We distinguish between datasets which were recorded using real devices and real networks, and those which were synthetically generated in the lab. The recorded datasets were created using standard commercial devices, networks, and protocols commonly used in Internet of Things applications aiming to resemble real world use cases. Several sessions were carried out to cover the influence of different parameters of payload and network types. The synthetically generated datasets simulate the influence of the network in a predefined way. All datasets were made open source and are available for download on our GitHub site[2].

### 4.1 Dataset Considerations

The recorded datasets (D-1 to D-5 and S-7 to S-10) were designed to resemble the behaviour and architecture of an Internet of Things use case, where many nodes are connected over a network. Each node continuously sends text-based messages to a common destination in a predefined interval over HTTP/1.1. The event producers (clients) were various kinds of Android smartphones, running a customized application which is optimized for efficient event generation. Two Windows PCs were also used as event producers for the WLAN datasets running the same code base. The sessions were recorded using either the internal wireless network after the IEEE 802.11 standard (WLAN), or the public cell phone network (UMTS) of different providers. Details of the hardware and software configuration of the used devices are provided on our GitHub site for the Datasets.

The synthetically generated datasets (G-1 to G-12) were created without a real network but instead by simulating different network behaviours in the computer. Twelve sessions were carried out with different network delay variations and also variations in the frequency of producing events. This results in datasets allowing us to get a better understanding of how dynamic buffer algorithms work and how to adapt their behaviour under various conditions.

A common way to produce a temporally correct ordered time-series event stream in a distributed system with independent event producers is to assign timestamps to events and to use this attribute to sort the event stream, requiring that the clocks of all nodes in the system are synchronized. The Network Time Protocol (NTP) [18] was evaluated for this purpose. An NTP client is already included in Android, but this operating system does not allow to intervene in the synchronization process without root permissions, which makes this variant useless for our approach. A simpler variant is the Simple Network Time Protocol (SNTP), but evaluations revealed that it is too imprecise for our use case. For this reason, we have implemented our own solution which is inspired by the synchronisation process of the Precision Time Protocol (PTP) [1]. Thereby the event producers' request the server time via HTTP/1.1 and then calculate the time difference to their internal clock. To get a properly synchronized clock, the actual implementation queries the server ten times and then calculates the median offset to the server time. This synchronization mechanism is automatically carried out before the start of each session.

[2]https://github.com/JR-DIGITAL/ooo-dataset

Temporal aspects play an important role in our work and therefore we collected several time-stamps to be able to fully reproduce the temporal occurrence of the event streams for later eval-uations. Figure4illustrates the whole chronological process starting from the event producer's detection of an event until it receives the response from the server. The following timestamps are involved:

- *Detection time ($dt$):* the time when the event producer (client) detects an event.
- *Client send time ($cst$):* the time when the message leaves the event producer.
- *Server receive time ($srect$):* the time when the server receives the event.
- *Server response time ($srest$):* the time when the internal processing of the server is finished and it sends the response to the event producer.
- *Client receive time ($crt$):* the time when the event producer receives the response from the server.

The following relevant durations can be derived from these timestamps:

- *Message preparation time:* the duration between the event producer's detection of an event and its sending of the message to the server ($cst - dt$).
- *Server processing time:* the duration the server needs to process the message ($srest - srect$).
- *Transmission time:* ($tt$) the duration between the event is detected until it reaches the server ($srect - dt$).
- *Network round-trip time:* ($RTT$) the duration where the message is on the network ($srect - cst$) $+(crt - srest)$.
- *Full processing time ($fpt$):* this includes the preparation time of the message, network round- trip time and server processing time ($crt - dt$).

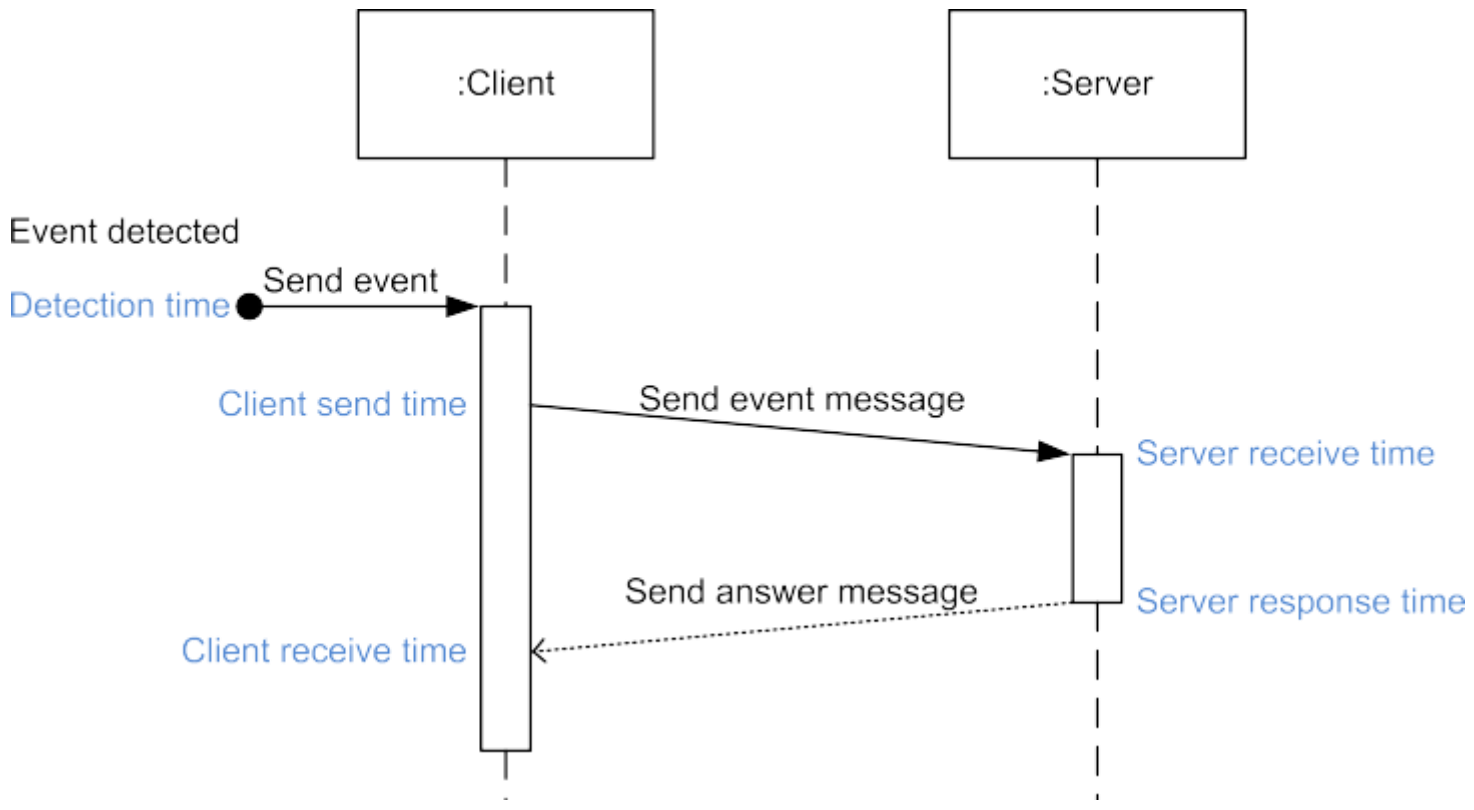

Fig. 4. Chronological sequence from detecting an event until the response is received by the event producer (client).

These datasets allow us to use two different ways to identify an event as an out-of-order event. The first is by using the sequence ID, which allows to identify out-of-order events per each event producer. Assuming we have an event stream of $e_1$, $e_2$, ..., $e_n$ to be correct when it is sorted in ascending order by the sequence ID $e_i.sid < e_{i+1}.sid$, $(1 \leq i \leq n)$, where $i$ is the order when the event was received in the fusion centre. We can identify an out-of-order event $e_j$ if there is an event $e_i$ with $e_i.sid > e_j.sid$. Although the fusion centre received it in the following order $1 \leq i < j \leq n$. The second approach is of greater interest in a distributed system with multiple event producers. A temporal property is used to reorder the event stream. Such a temporal property can be the detection time $dt$ which is generated by the event producer. Therefore we assume an event stream $e_1, e_2, ..., e_n$, $(1 \leq i < n)$ to be correct, when it is sorted in ascending order by the detection time $e_i.dt \leq e_{i+1}.dt$, where $i$ denotes the order when the event was received in the fusion centre. An out-of-order event $e_j$ can be identified if there is an event $e_i$ with $e_i.dt > e_j.dt$. Although the fusion centre received it in the following order $1 \leq i < j \leq n$.

## 4.2 Analysis

An overview of all recorded datasets is given in this document in tables 1 and 2 and the details can be found in the supplemental online document. The datasets were recorded in 21 sessions, with each session lasting 600 seconds. The datasets D-1 to D-5 have been done over the public cell phone networks (UMTS) of different providers with 7–9 clients. In these datasets, the clients sent events to the server in intervals of 500 ms. In each session we used a different predefined net payload between 0 bytes and 10 KiB. This results in a bandwidth for each client between 0.5 KiB/s and 21.4 KiB/s, and a bandwidth on the server between 4.1 KiB/s and 150 KiB/s. The out-of-order events, detected by using the detection time, range between 16.08 % and 34.14 % of the total events for each dataset.

For the sessions S-7 to S-10 we used our local WLAN during working hours. The interval time between events was set to 200 ms. This results in a bandwidth for the clients between 7 KiB/s and 53 KiB/s, and for the server between 69 KiB/s and 534 KiB/s. Detected out-of-order events range between 19.69 % and 28.32 % of the total events for each dataset.

The synthetically generated datasets (G-1 to G-12) were created using different functions for varying the network delay within a given range. Also the frequency of producing events varied between the datasets. All datasets were created on a single computer where the functions were simulated. No payload was added to these datasets as the network delays were simulated, and a payload would have no influence there anyway. The client side data rate for the datasets with a constant event producing frequency (datasets G-1 to G-8) is 1.3 KiB/s and 13 KiB/s for the server. In the datasets G-9 to G-12 the data rate varies on the client side between 0.03 KiB/s and 2.6 KiB/s, and on the server side between 0.3 KiB/s and 26 KiB/s. The number of out-of-order events in the G datasets ranges between 18.16 % and 81 % of the total number of events per dataset. The higher the bandwidth of simulated network delays, the higher are the detected out-of-order events.

The amount of out-of-orders events in the WLAN datasets (S-8 to S-10) is always higher than the amount in UMTS datasets (D-1 to D-5). This might be because of the lower interval of 200 ms of sending events in the WLAN dataset. The median and mean of the full processing time ($crt - dt$) in the UMTS datasets is always higher than the median and mean in the WLAN dataset. Figures of each dataset can be found in the online supplemental document which gives more insight into the behaviour over the session's duration. The transmission times for each event of a session including the detected out-of-order events are illustrated, as well as the frequency of event and out-of-order event occurrences.

# 5 ALGORITHMS

Seven different algorithms for out-of-order event compensation have been implemented in Java. Six of them dynamically recalculate the buffer time and one uses a static buffer which is used for comparison purposes. These dynamic buffer size algorithms are novel as, to the best of our knowledge, we are not aware that they were already presented elsewhere. All algorithms use the detection time ($dt$) to identify out-of-order events, as this is a suitable solution with multiple event producers in a distributed system. Incoming events are kept in the buffer until $dt$ *buffertime* is reached and are emitted after this period. If the buffer time is too small to correctly reorder an event, then it is marked as uncompensated and will be emitted immediately. The dynamic buffer algorithms continuously recalculate the buffer size based on the transmission times of the incoming events. The aim of a dynamic buffer is to adapt its buffer size according to the current environmental situation e.g. the varying transmission delays. This should allow to keep the buffer time as small as possible while reordering all incoming events. Subsequently we discuss the buffer size calculation of each proposed algorithm.

Table 1. An overview of the recorded datasets describing the number of clients, used network, the payload, and the resulting data rates.

| ID | Clients | Network | Interval (ms) | Net Payload (Bytes) | Gross Payload (Bytes) | Events | Server KiB/sec | Clients KiB/sec |
|---|---|---|---|---|---|---|---|---|
| D-1 | 8 | UMTS | 500 | 0 | 265 | 9600 | 4.14 | 0.52 |
| D-2 | 9 | UMTS | 500 | 512 | 1409 | 10800 | 24.77 | 2.75 |
| D-3 | 8 | UMTS | 500 | 1024 | 1365 | 9600 | 21.33 | 2.67 |
| D-4 | 7 | UMTS | 500 | 2048 | 2426 | 8400 | 33.17 | 4.74 |
| D-5 | 7 | UMTS | 500 | 10240 | 10929 | 8400 | 149.42 | 21.35 |
| S-7 | 10 | WLAN | 200 | 512 | 1409 | 30000 | 68.80 | 6.88 |
| S-8 | 10 | WLAN | 200 | 1024 | 1365 | 30000 | 66.65 | 6.67 |
| S-9 | 10 | WLAN | 200 | 2048 | 2426 | 29915 | 118.13 | 11.85 |
| S-10 | 10 | WLAN | 200 | 10240 | 10929 | 29999 | 533.64 | 53.36 |
| G-1 | 10 | Simulated | 200 | 0 | 265 | 30000 | 12.94 | 1.29 |
| G-2 | 10 | Simulated | 200 | 0 | 265 | 30000 | 12.94 | 1.29 |
| G-3 | 10 | Simulated | 200 | 0 | 265 | 30000 | 12.94 | 1.29 |
| G-4 | 10 | Simulated | 200 | 0 | 265 | 30000 | 12.94 | 1.29 |
| G-5 | 10 | Simulated | 200 | 0 | 265 | 30000 | 12.94 | 1.29 |
| G-6 | 10 | Simulated | 200 | 0 | 265 | 30000 | 12.94 | 1.29 |
| G-7 | 10 | Simulated | 200 | 0 | 265 | 30000 | 12.94 | 1.29 |
| G-8 | 10 | Simulated | 200 | 0 | 265 | 30000 | 12.94 | 1.29 |
| G-9 | 10 | Simulated | 10000 - 100 | 0 | 265 | 30057 | 0.26 - 25.88 | 0.026 - 2.59 |
| G-10 | 10 | Simulated | 10000 - 100 | 0 | 265 | 30058 | 0.26 - 25.88 | 0.026 - 2.59 |
| G-11 | 10 | Simulated | 100 - 10000 | 0 | 265 | 29502 | 25.88 - 0.26 | 2.59 - 0.026 |
| G-12 | 10 | Simulated | 100 - 10000 | 0 | 265 | 29503 | 25.88 - 0.26 | 2.59 - 0.026 |

Table 2. The analysis of the recorded datasets describing the number of out-of-order events and a summary of the processing times.

| | | | OoO Events | | Full Processing Time | | | | | | |
|---|---|---|---|---|---|---|---|---|---|---|---|
| ID | Clients | Network | Number | Percentage | Min | Q1 | Median | Mean | Q3 | Max | Std Dev |
| D-1 | 8 | UMTS | 1544 | 16.08% | 59 | 139 | 162 | 181.1 | 190 | 4738 | 103.9 |
| D-2 | 9 | UMTS | 3666 | 33.94% | 81 | 137 | 167 | 185.8 | 205 | 3680 | 115.6 |
| D-3 | 8 | UMTS | 3277 | 34.14% | 74 | 132 | 157 | 182.7 | 187 | 5616 | 183.8 |
| D-4 | 7 | UMTS | 2302 | 27.40% | 77 | 145 | 165 | 193.4 | 203 | 3300 | 116.5 |
| D-5 | 7 | UMTS | 1584 | 18.86% | 154 | 250 | 271 | 288.9 | 304 | 1911 | 83.7 |
| S-7 | 10 | WLAN | 7535 | 25.12% | 11 | 24 | 32 | 46.5 | 50 | 1522 | 49.9 |
| S-8 | 10 | WLAN | 5908 | 19.69% | 15 | 27 | 37 | 48.4 | 50 | 872 | 47.1 |
| S-9 | 10 | WLAN | 7880 | 26.34% | 15 | 29 | 39 | 52.6 | 54 | 3385 | 94.1 |
| S-10 | 10 | WLAN | 8495 | 28.32% | 46 | 75 | 92 | 103.7 | 114 | 1379 | 51.3 |
| G-1 | 10 | Simulated | 23968 | 79.89% | 101 | 302 | 503 | 500.9 | 699 | 901 | 229.7 |
| G-2 | 10 | Simulated | 5449 | 18.16% | 476 | 488 | 500 | 500.5 | 513 | 527 | 14.4 |
| G-3 | 10 | Simulated | 6328 | 21.09% | 29 | 300 | 526 | 525.7 | 751 | 1018 | 260.4 |
| G-4 | 10 | Simulated | 9034 | 30.11% | 24 | 301 | 526 | 525.4 | 750 | 999 | 260.2 |
| G-5 | 10 | Simulated | 20305 | 67.68% | 22 | 128 | 243 | 300.6 | 427 | 995 | 208.5 |
| G-6 | 10 | Simulated | 21099 | 70.33% | 22 | 129 | 244 | 300.8 | 429 | 991 | 209.0 |
| G-7 | 10 | Simulated | 8498 | 28.33% | 251 | 275 | 691 | 500.6 | 726 | 752 | 225.6 |
| G-8 | 10 | Simulated | 6652 | 22.17% | 251 | 342 | 500 | 500.6 | 659 | 750 | 159.8 |
| G-9 | 10 | Simulated | 24347 | 81.00% | 101 | 302 | 500 | 500.9 | 701 | 902 | 230.9 |
| G-10 | 10 | Simulated | 8096 | 26.93% | 448 | 488 | 501 | 500.7 | 513 | 554 | 14.4 |
| G-11 | 10 | Simulated | 23764 | 80.55% | 81 | 301 | 501 | 501.1 | 702 | 901 | 230.7 |
| G-12 | 10 | Simulated | 10406 | 35.27% | 447 | 488 | 501 | 500.6 | 513 | 554 | 14.5 |

### 5.1 Static Buffer Algorithm (SBA)

This algorithm uses a static, predefined buffer time. The Static Buffer Algorithm is included for comparison purposes to be able to evaluate the differences to other algorithms.

### 5.2 Buffer Sizing based on single Transmission Time (BSTT)

A dynamic value of the buffer time is necessary to achieve a better performance even when there are several changes to the network delay. Therefore, this algorithm uses the transmission time ($tt$) of the latest event and adapts the buffer size with an increase factor or decrease factor. The transmission time is the duration when the event was detected until it was received in the fusion centre. The *increaseFactor* and *decreaseFactor* define to which extent the buffer size will be changed. Additionally an offset is used which is a constant value and defines a safety margin for the buffer size. And the threshold, which is also a constant, damps the reduction of the buffer size. The algorithm works as follows: if $tt + offset$ is greater than the $currentBufferTime$, then the buffer time will be increased: $newBufferTime = (currentBufferTime * increaseFactor) + offset$. Respectively, if $tt + offset$ is smaller than the $currentBufferTime + threshold$, then the buffer size will be decreased as follows: $newBufferTime = currentBufferTime * decreaseFactor$. Thereby the reduction of the buffer size is damped by the threshold.

### 5.3 Buffer Sizing based on Transmission Time Weighted Average (BSTTWA)

This algorithm also uses a sliding window of predefined length containing $n$ transmission times to calculate a baseline of the overall network delay. In this case we use a weighted mean with exponentially decreasing weights and add an offset.

$$bufferTime = \frac{\sum_{i=1}^{n} (tt_i * w_i)}{\sum_{i=1}^{n} w} + offset \tag{1}$$

$$w_i = \left(\frac{n-i}{n}\right)^2 \tag{2}$$

### 5.4 Buffer Sizing based on Transmission Time Difference (BSTTD)

In the best case, such a buffer has to compensate only the variation of changes in the environment over time. Assuming that all delays were of constant length, there would be no out-of-order event. However, those delays are not guaranteed, especially in wireless networks or in networks with a shared medium. Therefore we calculate the maximum difference of transmission times over a sliding window of a predefined length and add an offset.

$$bufferTime = (max\,(tt) - min(tt)) + offset \tag{3}$$

### 5.5 Buffer Sizing based on Transmission Time Difference and Average (BSTTDA)

This buffer algorithm combines the calculation of an average with the transmission time difference. The average of transmission times gives a longer term baseline measure of the delays, and the transmission time difference adds the currently known delay variance. The calculations are done over a sliding window of predefined length, and an offset is added to these values as a safety margin.

$$bufferTime = \frac{1}{n}\sum_{i=1}^{n} tt_i + (max(tt) - min(tt)) + offset \quad (4)$$

### 5.6 Buffer Sizing based on Kalman Filter (BSKF)

The Kalman filter [24] is an algorithm which is able to predict the next value of a measurement through a given series of previous measurements. It estimates parameters of interest based on indirect, inaccurate and uncertain observations such as measurements from sensors. This is commonly used e.g. to track objects based on noisy data. The process model of a Kalman filter uses the combination of two equations:

$$X_k = A_{k-1} \cdot X_{k-1} + B_{k-1} \cdot u_{k-1} + w_{k-1} \quad (5)$$

where $X$ is the state vector with the variables of the system, $A$ is the matrix that defines the system,
$B$ is the control vector for each input in $u$ vector, and $w$ is the noise model vector. The second part, which defines the algorithm, is the following equation:

$$Z_k = H_k \cdot X_k + v_k \quad (6)$$

where $Z$ is the measurement prediction, $H$ is the transformation matrix, and $v$ is the noise measure- ment vector. In this buffer algorithm, the Kalman filter is used to predict the next transmission time ($tt_{predicted}$), based on previous observations. An offset is added to the predicted value as a safety margin:

$$bufferTime = tt_{predicted} + offset \quad (7)$$

For the realisation of this algorithm we use the implementation of the Kalman filter of the Apache Commons Math project[3].

### 5.7 Dynamic K-Slack Buffer Sizing (KSLACK)

The original K-Slack algorithm [3] did not define a dynamic adaptation of the buffer size. Later a dynamic adaptation of $K$ was proposed in [19]. This dynamic buffer sizing algorithm was imple- mented for the comparison of our proposed algorithms. We omitted the derivation of the clock from incoming events since we already have a synchronized and stable clock. The buffer size (in the paper referenced as $K_d$) is the maximum transmission time of all observations plus an offset. The offset is an added safety margin which is the product of the standard deviation $\sigma$ of all transmission delays and a scaling factor $\lambda$. The scaling factor must be defined in advance.

$$bufferTime = max[e.tt] + (\sigma * \lambda) \quad (8)$$

$$\sigma = \sqrt{\frac{1}{n-1}\sum_{i=1}^{n}(e_i.tt - \overline{e.tt})^2} \quad (9)$$

### 5.8 Experimental Setup

To find the optimal parameters for a buffer sizing algorithm is a difficult task as each algorithm has different parameters and each parameter influences the behaviour in a different way. Figure 5 illustrates this exemplary with the algorithms SBA, BSTTWA, and BSTTDA on dataset S. An evaluation of all parameters for all algorithms can be found in the supplemental online document.

For a set of initial values for the parameters we suggest to record a sample data set with around 5,000 events. Then calculate the *full processing duration* for all events. This is the duration when the

[3]http://commons.apache.org/proper/commons-math/

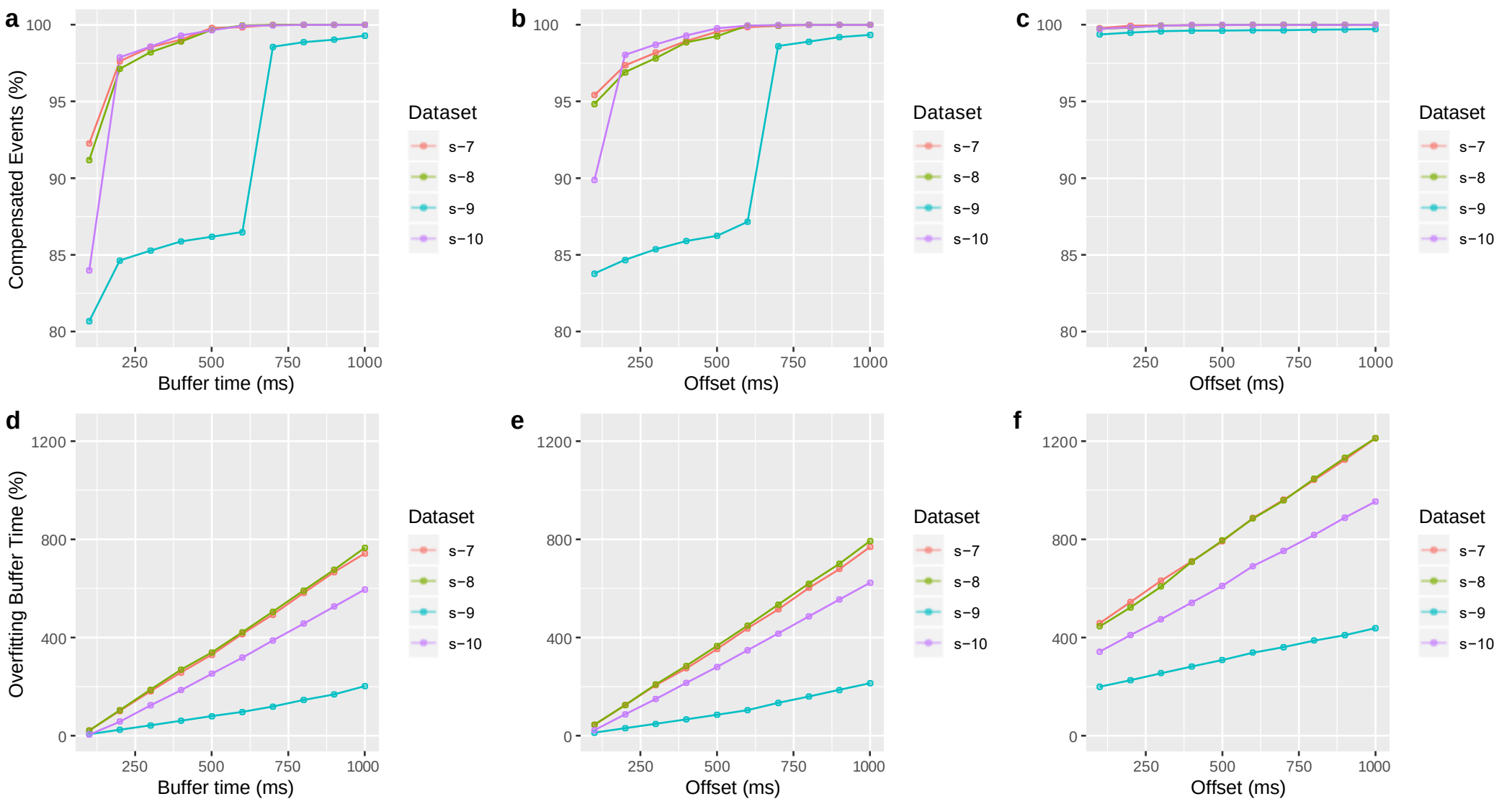

Fig. 5. Influence of algorithm parameters on the resulting compensated out of order events and the buffer time for dataset S. Algorithm SBA with the parameter “buffer time” (a, d), algorithm BSTTWA with the parameter “offset” (b, e), and algorithm BSTTDA with the parameter “offset” (c, f).

event occurred in the real world until it was received at the processing node which fuses events (event receive time - event occurrence time). Based on this we suggest the following starting values:

For all algorithms: the *initial buffer time* is mainly to avoid the cold start problem and it can be set to: $98^{th}$ percentile of full processing duration * 5

- *Static Buffer Algorithm (SBA)*
  buffer time: $98^{th}$ percentile of full processing duration * 6
- *Buffer Sizing based on single Transmission Time (BSTT)*
  offset time: $98^{th}$ percentile of full processing duration * 3
  threshold to decrease the buffer 100 ms
  increase factor: 2
  decrease factor: 0.99
- *Buffer Sizing based on Transmission Time Weighted Average (BSTTWA)*
  number of samples (size of the sliding window): 100;
  offset time: $98^{th}$ percentile of full processing duration * 4
- *Buffer Sizing based on Transmission Time Difference (BSTTD)*
  number of samples (size of the sliding window): 600;
  offset time: $98^{th}$ percentile of full processing duration * 2
- *Buffer Sizing based on Transmission Time Difference and Average (BSTTDA)*
  number of samples (size of the sliding window): 600;
  offset time: $98^{th}$ percentile of full processing duration
- *Buffer Sizing based on Kalman Filter (BSKF)*
  offset time: $98^{th}$ percentile of full processing duration * 4
  Turn off the control and noise vectors.

- *Dynamic K-Slack Buffer Sizing (KSLACK)*
  scaling factor: 0.8
  initial buffer time: $98^{th}$ percentile of full processing duration * 4

## 6 EVALUATION AND RESULTS

The evaluation of the algorithms was carried out with all datasets resulting in 147 individual runs. The summarized results of all runs are visualized in figures 6, 7 and 8. Each individual run of an algorithm over a dataset is visualized in detail in the supplemental online document. Based on the parameter evaluation described above, each algorithm was configured individually and the configuration was kept the same for each run. The following settings where applied to the algorithms:

- *Static Buffer Algorithm (SBA)*
  buffer time: 1000 ms
- *Buffer Sizing based on single Transmission Time (BSTT)*
  initial buffer time: 500 ms; threshold to decrease the buffer 100 ms; increase factor: 2; decrease factor: 0.99; offset time: 500 ms
- *Buffer Sizing based on Transmission Time Weighted Average (BSTTWA)*
  number of samples (size of the sliding window): 100; initial buffer time: 750 ms; offset time: 750 ms
- *Buffer Sizing based on Transmission Time Difference (BSTTD)*
  number of samples (size of the sliding window): 600; initial buffer time: 750 ms; offset time: 350 ms
- *Buffer Sizing based on Transmission Time Difference and Average (BSTTDA)*
  number of samples (size of the sliding window): 600; initial buffer time: 750 ms; offset time: 350 ms
- *Buffer Sizing based on Kalman Filter (BSKF)*
  initial buffer time: 750 ms; offset time: 600 ms. The control and noise vectors were disabled.
- *Dynamic K-Slack Buffer Sizing (KSLACK)*
  scaling factor: 0.8; initial buffer time: 750 ms.

Dynamic buffer sizing algorithms need to keep the buffer size low while compensating all out-of-order events. Following metrics were used in the evaluation which reflect these characteristics.

- *Not compensated events (%)*: is the percentage ratio of events which the algorithm did not compensate to all out of order events in a dataset.

$$notCompensatedEventsPercentage = \frac{notCompensatedEvents}{allOutOfOrderEvents} * 100 \qquad (10)$$

- *Overfitting buffer time (%)*: is the percentage ratio of actual buffer size to the minimal required buffer size to be able to compensate all events. The overfitting of the buffer time is caused by the algorithms added safety margin.

$$overfittingBufferTimePercentage = \frac{bufferTime}{minimumBufferTime} * 100 \qquad (11)$$

The static buffer algorithm (SBA) is the baseline for all other algorithms. A fixed buffer size of 1000 ms was used. It can achieve good results when the buffer size is set high enough. This is visible in the results of the synthetic datasets G where it produces throughout good results. But on real world datasets the static buffer fails, especially on the datasets D-4 and D-5 where it results in a higher number of not compensated events.

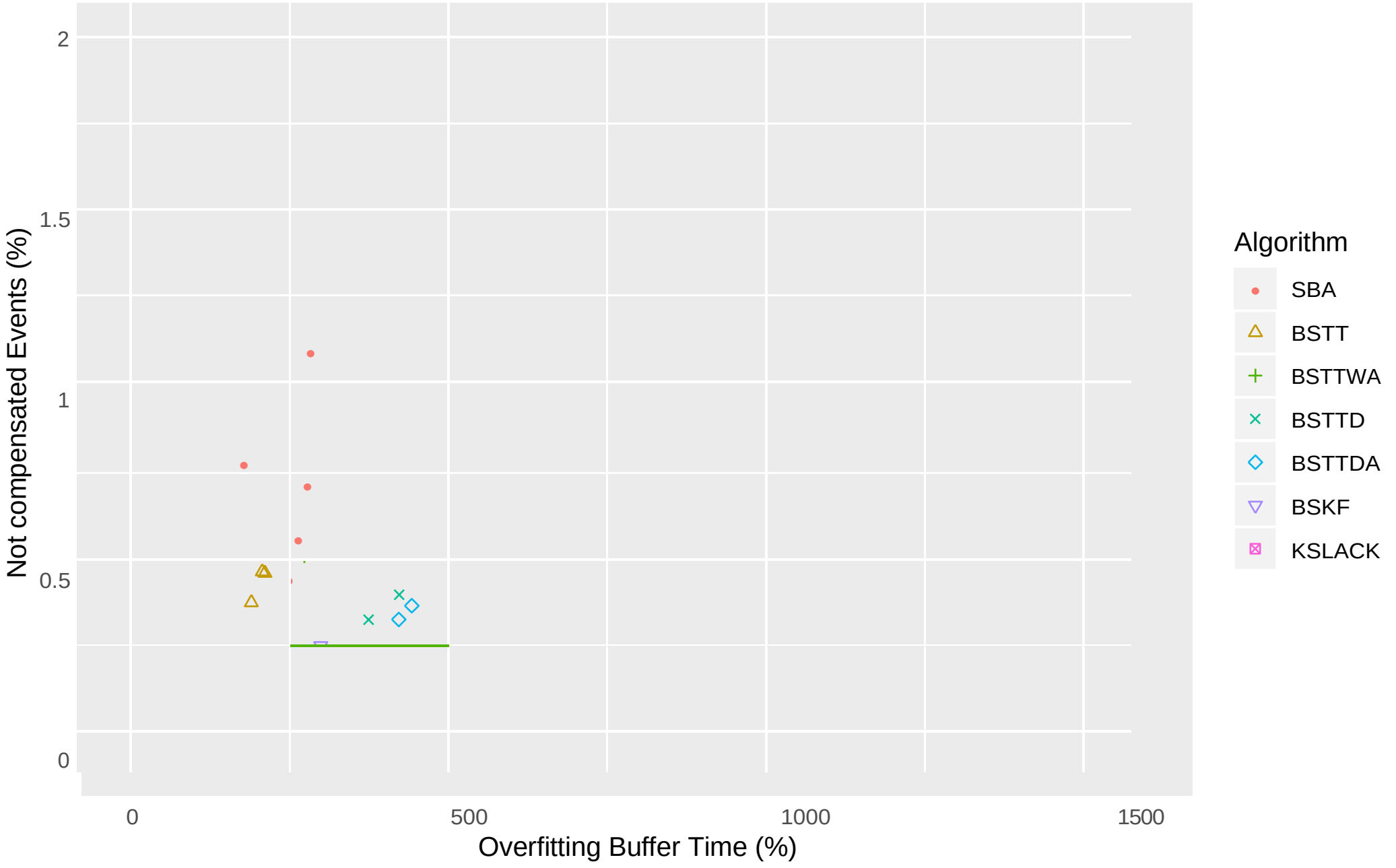

Fig. 6. The result of each algorithm for the dataset D with overfitting buffer time on the x-axis and not compensated events on the y-axis. Lower values indicate better results.

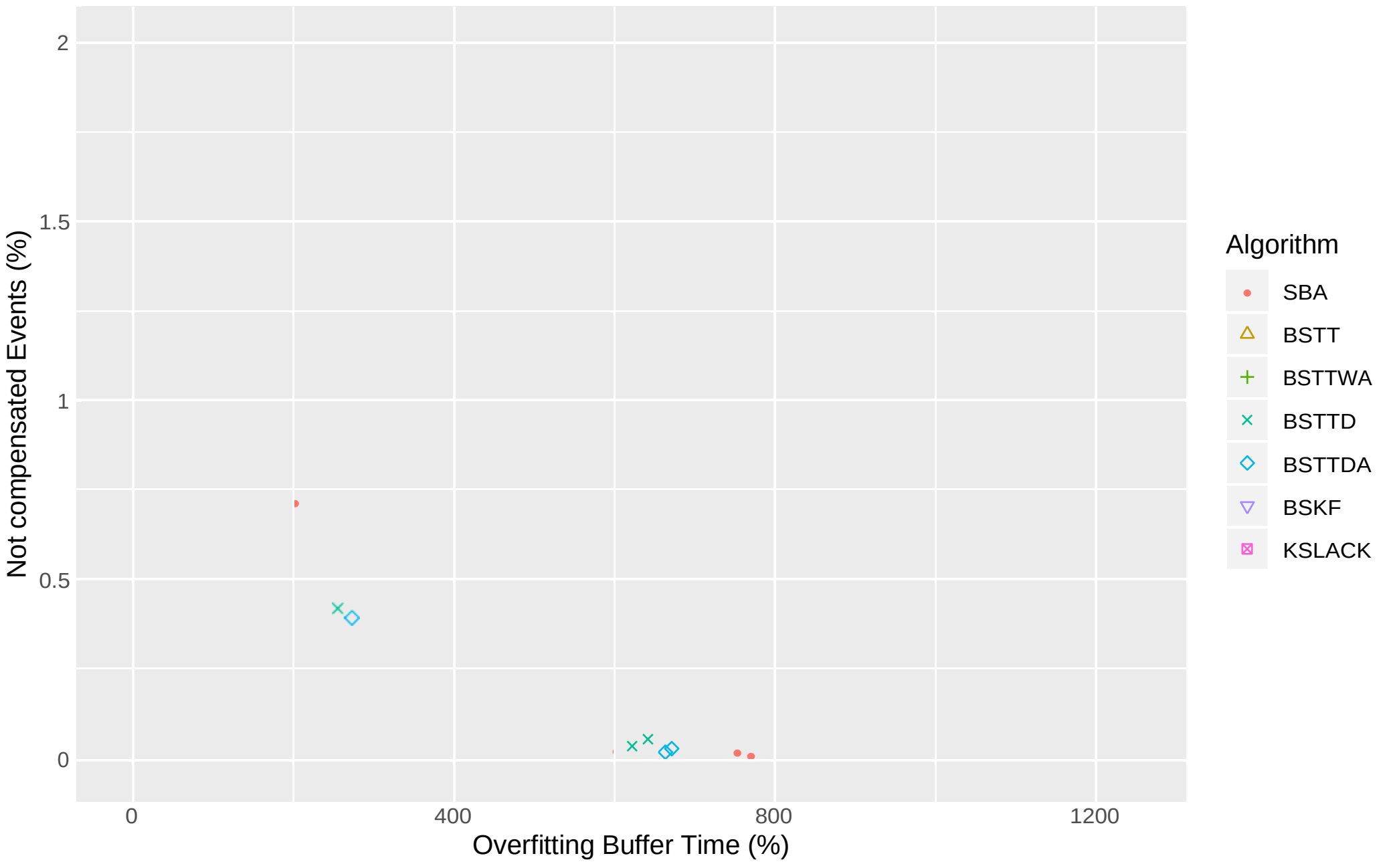

Fig. 7. The result of each algorithm for the dataset S with overfitting buffer time on the x-axis and not compensated events on the y-axis. Lower values indicate better results.

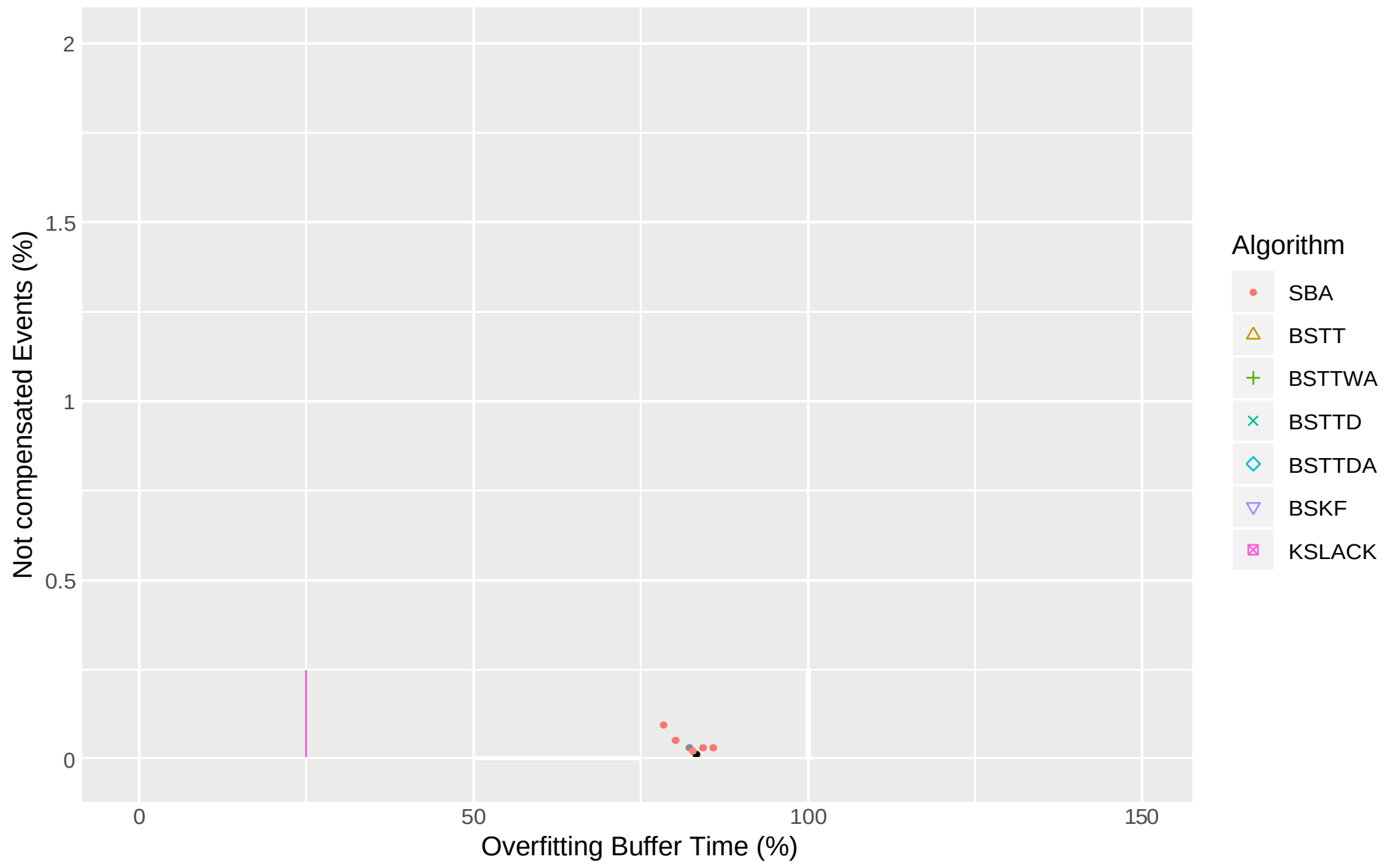

Fig. 8. The result of each algorithm for the dataset G with overfitting buffer time on the x-axis and not compensated events on the y-axis. Lower values indicate better results.

The algorithm "buffer sizing based on single transmission time (BSTT)" uses the transmission time of the latest event to calculate the optimal buffer size. It performed well on dataset D-5, but failed on the S datasets, where it produced the highest number of not compensated events in 3 out of 4 cases. It produced good results on the G datasets overall, meaning it can handle various situations with the exception of a decrease in event occurrence frequency (see also the result on dataset G-11). The variance of the buffer size is much higher when the variance of the transmission times in the dataset is high (see also figure9(e)). For example, using the dataset dataset G-1 (dataset full processing time ($fpt$) std dev. 229.7) this algorithm has a standard deviation of the buffer time of 58.7, in comparison to the algorithm BSTTD which had a standard deviation of the buffer size of
12.7 in this case. A disadvantage of this algorithm is, that it requires various parameters to be set. On the other hand, it does not use a sliding window and it is not as badly affected by the cold start problem as other algorithms.

A weighted average over a sliding window, which gives newly arrived events a higher importance was the idea of the BSTTWA algorithm. This algorithm achieved good results on the datasets S-7, S-8, and S-10 in comparison to other algorithms on this dataset. A drawback of this approach is, that it cannot adapt to sudden and big changes of transmission times of incoming events, as illustrated in figure9(a). Applied on the datasets G, it is visible that it can handle various changes in the transmission times, except in the case of dataset G-11. An increase of transmission time variance could also pose a problem. As visible in the buffer size behaviour in dataset G-5 and G-6, the difference between buffer size and maximum transmission time decreases when the transmission time variance gets higher. The transmission time's mean over a sliding window provides a good overall measure of the current state of the network. BSTTWA produced useful results on the datasets D and S, but it requires an overall higher offset value.

To calculate the buffer size, the BSTTD algorithm uses the difference between the minimal and maximal transmission times over a sliding window, as, ideally, all you have to compensate for is the variance of the transmission time. This algorithm needs a fairly big sliding window to work reliably and hence suffers especially from the cold start problem. In comparison to other algorithms the BSTTD implementation achieved throughout good results on the datasets D and S. This algorithm is perfectly equipped to adapt the buffer size to increasing or decreasing transmission time variances (see also the results on dataset G-5 and G-6). But at the same time, it fails to adapt the buffer size when the transmission time increases and the variance stays the same (see also the results on dataset G-3, G-4, and9(c)). It has the ability to adapt to sudden changes (see also the results on dataset S-9) and - in comparison to other algorithms - it needs only a fairly small offset value of 350 ms.

A Kalman filter is often used to process noisy sensor data, and therefore we use this approach as the basis to calculate a suitable buffer size in the BSKF implementation. In the evaluation experiments, the performance of the BSKF algorithm is below average. The BSKF algorithm also has a problem when transmission time variances increase, visible on datasets G-5 and G-6. One positive aspect is that it suffers less from a cold start problem in comparison to competitors which use a sliding window.

For the comparison of state-of-the-art dynamic buffer sizing algorithms with our proposed algorithms, the dynamic buffer sizing for K-Slack algorithms, as described in [19], was implemented. This algorithm is easy to configure, as it just needs a scaling factor. For this, some runs with different values were carried out, and the best option was chosen. This K-Slack algorithm produced on the Datasets D and S throughout a low number of not compensated events, but at the same time the buffer size was nearly the highest on all datasets. One big disadvantage is, that it does not sufficiently decrease the buffer size when the transmission time of the data set decreases. This gets visible on the synthetic datasets G (see also figure9(f)), and it is an indicator that this algorithm is not a good choice when a low buffer size is desired.

The idea of the BSTTDA algorithm is to combine the features of the algorithm BSTTD with an average value. An average over a sliding window provides a good overall measure of the current transmission time, and then adding the maximum difference of transmission times should also allow this algorithm to adapt to sudden changes of transmission times. As illustrated in9(b), this is the case. This algorithm can also handle increasing or decreasing transmission times while the variance of the transmission times is consistent, see also the results on datasets G-3 and G-4. A notable characteristic is that it overcompensates increasing transmission time variances, see the results on datasets G-5 and G-6. Though this algorithm suffers from a cold start problem, it achieved throughout good results. An advantage is also that it needs a fairly small offset value of 350 ms (same as for BSTTD). The influence of the size of the sliding window is negligibly small.

We deliberately set the offset values for all algorithms as low as possible to make the limitations of each algorithm clear. Hence, the key for reordering all out-of-order events in all possible situations is to give the algorithms enough offset, but this results in high buffer times which might be unwanted. For situations where the reaction time is more important than compensating all out-of-order events, the BSTTWA algorithm might be the choice. This algorithm is stable over a long time, neglects single outliers but still adapts to changes. The advantages of the BSTTDA algorithm are that it can handle various situations such as changing transmission times and changing transmission time variances as well as respond to sudden changes in transmission times, while needing only a small offset value (see also figures9(b) and (d)). These characteristics make this algorithm a good choice for most applications.

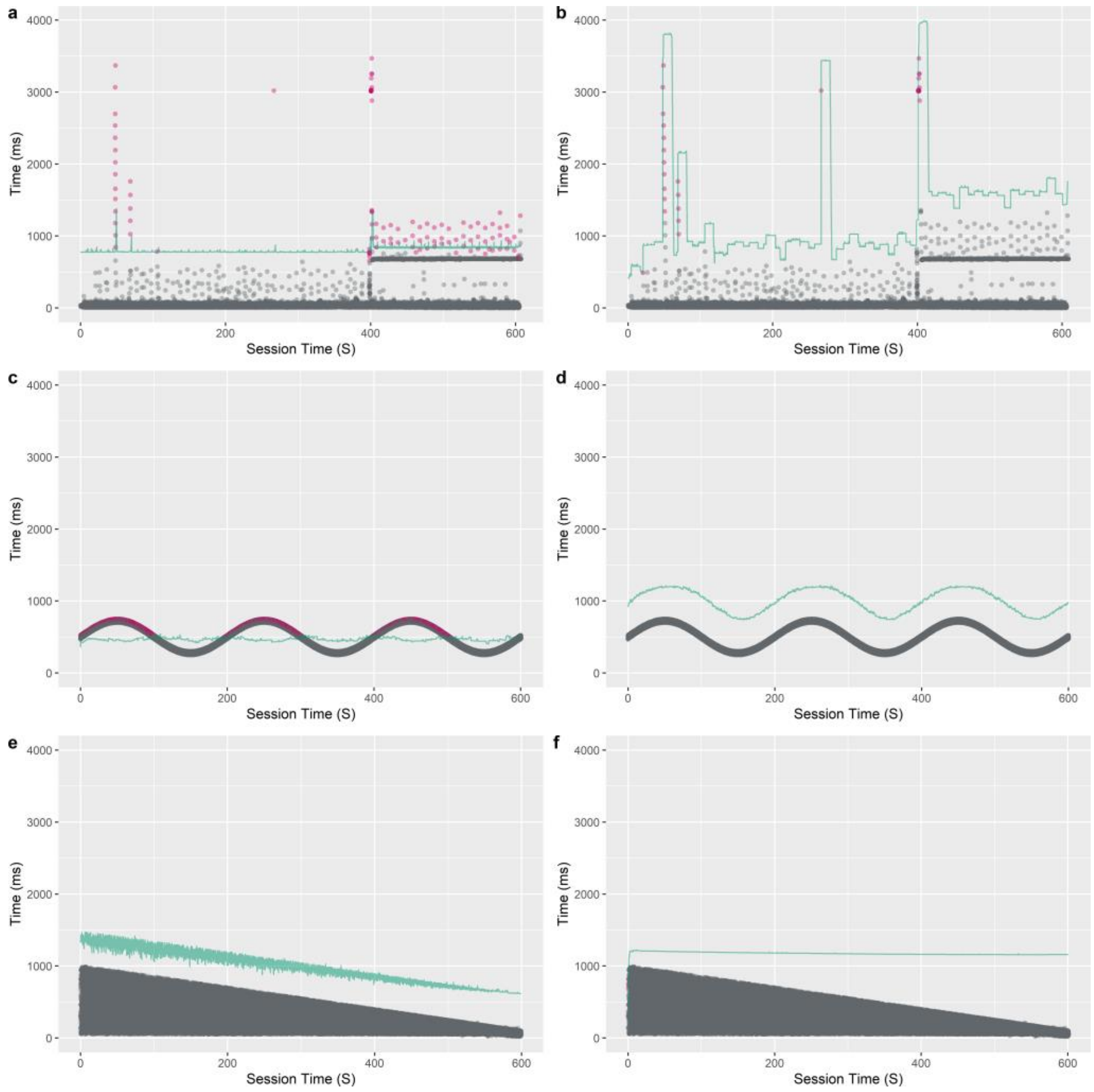

Fig. 9. Examples of dynamic buffer sizes: these figures illustrate the buffer size of the algorithm (green line) and the transmission times for each event represented as a point. Uncompensated out-of-order events are highlighted in pink. (a) algorithm BSTTWA on dataset S-9 (failed to adapt to environmental changes), (b) algorithm BSTTDA on dataset S-9 (better adaptation to environmental changes), (c) algorithm BSTTD on dataset G-8 (failed to adapt to biased sine wave with smaller delay variance), (d) algorithm BSTTDA on dataset G-8 (good adaptation to biased sine wave with smaller delay variance), (e) algorithm BSTT on dataset G-6 (good adaptability of the buffer size, but the buffer time variance correlates with the dataset variance), (f) algorithm KSLACK on dataset G-6 (does not sufficiently decrease the buffer size when the transmission time of the dataset decreases).

## 7 DISCUSSION

The original idea of this project was to find dynamic buffer sizing algorithms for multiple, distributed and independent sources, which allow subsequent time-sensitive applications to work correctly. The first question was if there is a general applicable event fusion mechanism for multiple and distributed sources dealing with temporally disordered event streams. We found various related research covering these topics. Much of this research is application-specific, has certain prerequisites to event producers e.g. that they know each other, or that knowledge of the system architecture is available. But this is not applicable in the case of an Internet of Things application where event producers can be attached and detached whenever necessary. What we need is an out-of-order event compensation approach which is independent from the source and sink.

The criteria when a dynamic time-out buffer method is preferable to a static buffer are: 1. that the overall delay is smaller while reordering more out-of-order events, and 2. that it is able to quickly adapt its buffer size to even sudden environmental changes, e.g. varying transmission delays. The evaluation has shown that the use of a dynamic time-out buffering method is to be preferred over a static buffer. The higher the variation of the network or other influences in the environment, the more necessary it is to use an algorithm which dynamically adapts its buffer size. A forecast for a suitable buffer size for a static buffer might be difficult and not always applicable. Based on our findings in the evaluation, we encourage to use the "Buffer Sizing based on Transmission Time Difference and Average (BSTTDA)" algorithm. It is also clear, that the algorithms BSTTWA, BSTTD, and BSTTDA outperform the current state-of-the-art algorithm for dynamic K-Slack buffer sizing (KSLACK).

We can also agree that a dynamic time-out buffering method is applicable and useful to fuse data from multiple distributed and independent sources for time-sensitive applications. The only prerequisite is that reasonably accurate and synchronized clocks are available.

One restriction still exists: we cannot guarantee full temporal order. On the one hand, this lies in the nature of distributed systems, where system crashes or network failures lead to independent failures. In such a case, parts of the system are isolated which might not be immediately visible to all other nodes. On the other hand, we have recognized in our recorded evaluation datasets that it is difficult to estimate the rare outliers with a high transmission time. The use of buffering algorithms is still a trade-off between reaction time and out-of-order event compensation, but it is useful in various applications. The varying transmission times of events in the recorded datasets could stem from bloated buffers in the network, as described by [10]. It should be considered to compensate these bloated buffers with additional, dynamically, and perhaps growing buffers.

In the introduction we listed two use cases where the usage of such fusion and compensation mechanisms is essential to guarantee a correctly working system. In general, such a fusion and compensation mechanism is a prerequisite when a temporally correct ordered event stream is necessary to guarantee correct results. It becomes more important when additionally one or more of the following aspects are true: the system is deployed in a distributed fashion, with many event producers, in distant locations, using low power devices, or networks with limited capacity, or with high a frequency of updates - just to name a few aspects. Then, the occurrence of out-of-order events is more likely. Such applications can be machine monitoring systems, Internet of Things applications, or sensor networks.

## 8 CONCLUSIONS

Event data fusion in sensor networks and in distributed systems in general is a requirement for various applications. When it comes to feed the data into time-sensitive applications, it becomes crucial that the fusion mechanism considers the temporal aspect of events in a sensible way. We presented two use cases from different domains where temporally correct ordered event streams are essential to guarantee correct results. The same holds true for a multitude of applications in the field of distributed systems, where it is likely that out-of-order events can occur. The reasons for out-of-order events are manifold. It could be the non-deterministic process scheduler in an operating system, the shared medium of a wireless network, or just the varying time which is needed to detect a certain event in the real world. In this paper we gave an introduction into this problem domain, looked at influencing characteristics of distributed systems, enumerated what can happen when out-of-order events occur, and discussed how to deal with them in general.

We were researching dynamic buffer sizing algorithms for multiple, distributed and independent sources, which reorder event streams so that subsequent time-sensitive applications work correctly. To be able to evaluate such algorithms we had to record datasets first, as we did not find a freely available dataset with all the desired features. Two types of datasets were created, one type which was recorded using real devices and real networks, while the other type of datasets was synthetically generated in the lab. These datasets were contributed to the scientific community, allowing them to evaluate and compare their compensation algorithms.

Five novel algorithms were implemented which dynamically adapt their buffer size based on different adaptation strategies. These algorithms were evaluated with a state-of-the-art approach for dynamic K-Slack adaptation, and with a static buffer. The implemented dynamic buffering algorithms are able to adapt their buffer sizes to environmental changes, such as varying transmission delays or other influences. The evaluation has shown that the use of a dynamic time-out buffering method is preferable to a static buffer. The higher the variation of the network or other influences in the environment, the more necessary it becomes to use an algorithm that dynamically adapts its buffer size. Based on our findings in the evaluation, we encourage to use the "Buffer Sizing based on Transmission Time Difference and Average (BSTTDA)" algorithm, as it produced overall good results in the evaluation. We could also show that this algorithm clearly outperforms the state- of-the-art approach for dynamic K-Slack buffer sizing. These types of algorithms are universally applicable, easy to integrate into existing architectures, and particularly interesting for distributed time-sensitive applications.

Time sensitive event data fusion for multiple, distributed and independent sources builds on time, and therefore clock synchronization becomes a crucial aspect on sensor networks. Using time-out methods allows for partial order guarantee, but consequently there is still a trade-off between reaction time and out-of-order event compensation. Ultimately, it depends on the application use case how much delay and how much out-of-order events are desired.